# INTELLIGENT TECHNOLOGIES IN MODEL BASE MANAGEMENT SYSTEM DESIGN AUTOMATION

## I.I. Semenova

*Russia, Omsk, Siberian state automobile and highway academy*
*osobaii@gmail.com*

The article describes the prospects of model base management system design automation for decision support systems and suggests the toolbox scheme for design automation based on intelligent technologies.

Knowledge representation models have become an integral part of the intelligent decision support system (IDSS). This area is covered in many different works. The subject of particular interest is the application of intelligent technologies for the development of the tools of decision support system (DSS) design automation and its distinct elements, namely model base management systems (MBMS).

Model base management system as a part of DSS includes the elements, as in figure 1 [2].

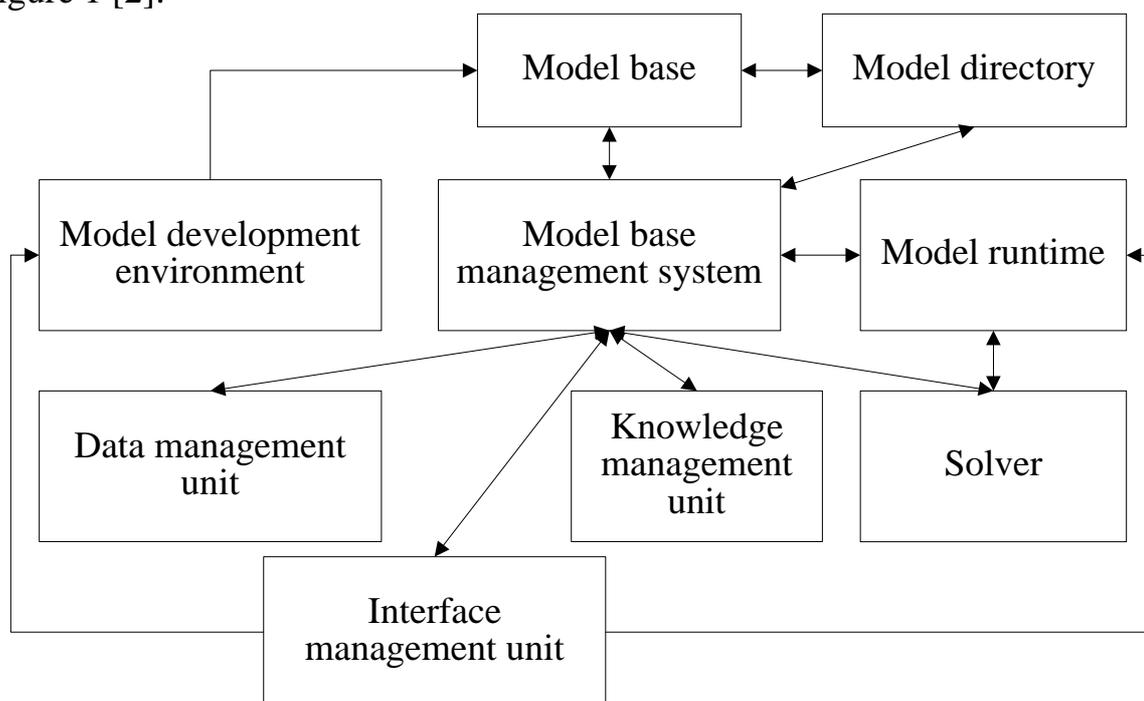

*Figure 1. Model base management system structure*

We use the system blocks description presented in [2].

***Model base*** is a set of computer decision models. Its functions are similar of those of data base, the only difference is that its stored objects are models. The models in the model base can be divided, in accordance with their type, into several categories such as: strategic, tactical, operational and analytical.

***Model base management system*** is a software package providing an access to a model base and its connections with other components. It often includes *model development environment* (MDE) and also *model runtimes* (MR) for model processing in a model base.

The function of the *model directory* equals to that of a data base directory. It is a catalogue of all models and other software in the model base. It has definitions of the models and their main functions to answer the questions of availability and compatibility of the models.

*Model development environment* (MDE) supports the process of building a model in order to achieve the highest level of usability. It should contain the *model definition language* (MDL), so that the models have been properly represented and stored in the model base for execution. It also provides the platform where models can be created, stored, integrated, selected and maintained when necessary.

*Model runtime* (MR) includes the *model manipulation language* (MML), which starts the execution of the existing models to get the best possible solution. It also contains the user's interface for managing the models that have been selected, and the links to solvers and data management modules.

**Solvers**, or solver-systems, are program tools that help users manage the models and find the solution (including the optimal one) for the problem by following a definite procedure. For example, the means of implementing linear programming techniques are described in [1].

The elements described above are connected with the corresponding data and *knowledge management subsystems* (fig. 1) and available to the persons making decisions (PMD) via DSS *user interface.*

The project of the model base management system is affected by the way of representing a compound hardware (CH), for which operation mode with the choice of the optimal or acceptable one for reaching the goals will be reproduced in MBMS. Besides, the structure is influenced the decision making methods and optimization methods a PMD intends to use in their work. A PMD's wish to modify the parameters of CH models and to compare the modeling results considering those parameters.

When designing model-oriented DSS, a client might happen to have existing software solutions which can do certain tasks in MBMS, for instance, they have a CAE system (e.g. AnyLogic, MatLab, MathCad etc.), solver or the systems executing certain solutions or optimization methods (e.g. BendX Stochastic Solver, LINDO API, OML, Risk Solver Platform etc.). Then it will be necessary to integrate existing solutions with the MBMS being designed.

According to the decision making scheme in DSS and the methods and tools helping to implement this decision, while automating MBMS design you need to make up a knowledge base containing the information about patterns and connections between decision making stages and the elements (units that are existing software modules with formally described structure for visualization and analysis of correctness of MBMS scheme building) that should be used in MBMS. Function diagram of means of MBMS automation design is shown in figure 2.

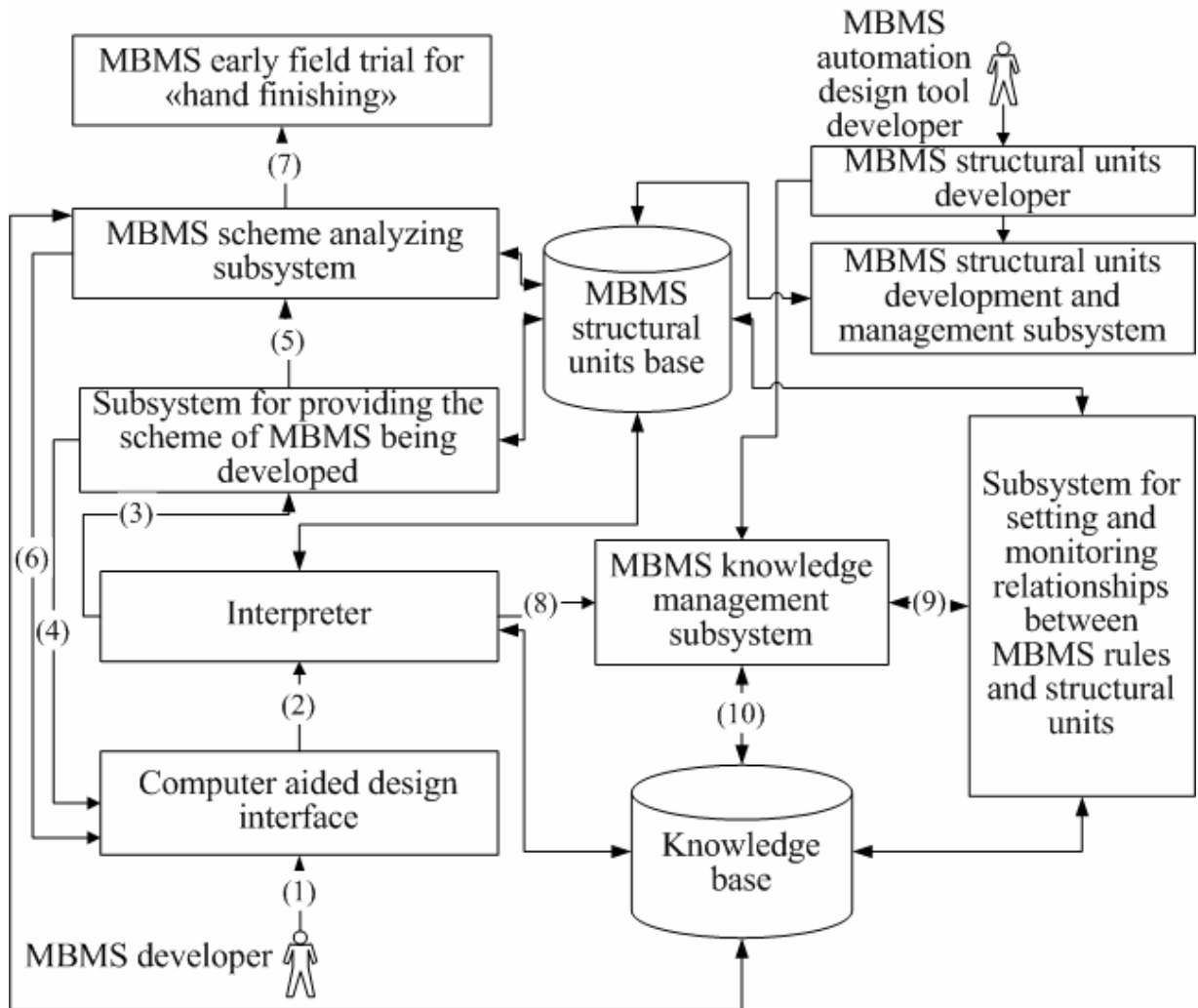

*Figure 2 — Organizational chart of means of creating MBMS automation design, where: (1) requirements for the system in development; (2) formalized requirement; (3) selected MBMS units and their integration into the general scheme in accordance with the rules; (4) request of the next requirement; (5) formalized description of the MBMS project; (6) revealed mistakes, recommendations for correction of the scheme; (7) generation of MBMS code; (8) absence of requirement processing rules; (9) connecting new rules with MBMS units; (10) adding new rules.*

When designing the knowledge base for MBMS design automation, the key points are the way of representing facts, patterns etc. and also the choice of storage structure. After analyzing the knowledge representation model classification, namely the peculiarities of the logic models (deductive, inductive, pseudophisical logics etc.) and heuristic models (functional networks, scripts, semantic networks, frames, productions etc.) a conclusion has been made that production and frame models can be selected as a knowledge representation model.

Production model enables to reflect the stored knowledge about MBMS structural parametric synthesis. Frame model is convenient to use when analyzing MBMS scheme (fig. 2) by comparing its current variant with prototype frame and / or pattern frame.

In order to maintain the means of MBMS automation project in working state it is necessary to provide the conditions for the knowledge base development. One of the solutions of this problem presumably can be found in creating a toolbox according to Open source methodology and an access to the system as a web-resource with an option of implementing the local version (to the client's wish) with inheritance of all the knowledge base or its part meeting the client's demands.

## References

1. Linear programming software survey// OR/MS Today Journal.–July 2009.– URL: http://www.lionhrtpub.com/orms/surveys/LP/LP-surveymain.html

2. Turban E. Decision Support and Business Intelligent System / E. Turban, J.E. Aronson, T.P. Liang, R. Sharda.– NJ: Prentice Hall, 2007.– 850 p.